\documentclass[%
 reprint,
 amsmath,amssymb,
 aps,
prb,
]{revtex4-2}

\usepackage{graphicx}% Include figure files
\usepackage{dcolumn}% Align table columns on decimal point
\usepackage{bm}% bold math
\usepackage{color}  
\begin{document}

\preprint{APS/123-QED}

\title{Topological states and flat bands induced by bound states in the continuum 
in a ladder-shaped one-dimensional photonic crystal}% Force line breaks with \\
%\thanks{A footnote to the article title}%

\author{Sofia Pinto}
 \email{sofia.pinto@sansano.usm.cl}
 \affiliation{Departamento de F\'isica, Universidad T\'ecnica Federico Santa
Mar\'ia, Casilla 110-V, Valpara\'iso, Chile.}
\affiliation{Instituto de Física, Pontificia Universidad Católica de Valparaíso, Av. Brasil 2950, Valparaíso, Chile.}
\author{P. A. Orellana}%
 \affiliation{Departamento de F\'isica, Universidad T\'ecnica Federico Santa
Mar\'ia, Casilla 110-V, Valpara\'iso, Chile.}
\author{Sergio Bravo}
 \affiliation{Departamento de F\'isica, Universidad T\'ecnica Federico Santa
Mar\'ia, Casilla 110-V, Valpara\'iso, Chile.} 

\date{\today}% It is always \today, today,
             %  but any date may be explicitly specified

\begin{abstract}

One-dimensional crystals serve as a versatile platform for engineering nontrivial states, which
can be easily explored in transport configurations. In this work, we analyze the properties
of a periodic structure composed of an H-shaped unit cell, which forms a periodic
ladder-shaped system. Using tight-binding models, group-theoretical considerations, and
standard band topology, we uncover the influence of bound states in the continuum (BICs) and
quasi-BICs formed in the original finite geometry on the creation of nontrivial band states.
By designing various textures for the onsite energies, we discovered a topological band
inversion between quasi-BIC-induced bands, leading to the emergence of topologically protected
edge states that are characterized by a quantized Zak phase. Additionally, we found an on-site
configuration that exhibits robust flat bands, induced by a symmetry-protected BIC and linked to
special one-sided localized edge states. We present a detailed analysis of the mechanisms driving
both effects and discuss the crucial role of symmetry in characterizing the topological phases of
these systems.

%\begin{description}
%\item[Usage]
%Secondary publications and information retrieval purposes.
%\item[Structure]
%You may use the \texttt{description} environment to structure your abstract;
%use the optional argument of the \verb+\item+ command to give the category
% of each item. 
%\end{description}
\end{abstract}

%\keywords{Suggested keywords}%Use showkeys class option if keyword
                              %display desired
\maketitle

%\tableofcontents

\section{\label{sec:level1}Introduction \protect} %\\ 
%break was forced \lowercase{via} \textbackslash\textbackslash}

%Photonic crystals introduction. Paraxial approximation, 
%1d crystals topology. BICs.\\
Controlling the propagation of light in structured materials has become a pivotal focus 
in modern photonics, with important implications for a variety of applications, including
integrated optical circuits and quantum technologies. In this context, photonic crystals
(PC), which were independently proposed by Yablonovitch \cite{Yablonovitch} and John
\cite{Jhon} in 1987, have emerged as a promising platform for light manipulation 
\cite{Gansch2016_Light}.
PC are structures where the dielectric function of the material varies periodically in
space, creating an analogy to electrons moving within a lattice potential. This periodic
modulation results in allowed and forbidden propagation bands, known as photonic band
gaps, which enable the control of light properties \cite{PC_book_prince}. These 
structures have aroused great interest in recent years due to the possibility of explore
new fundamental physics effects \cite{Lustig2023_PtimeC,PLR2025_PC_newphysics} and also a
wide variety of technological applications
\cite{BUTT2021_OPtics_PC_rev1,sciadv2025_topPC_fiber,2025_PC_sensors_rev}.

In parallel with the development of PC and their associated topological properties,
another phenomenon of great scientific interests are bound states in the continuum (BICs).
Originally proposed by von Neumann and Wigner in 1929 \cite{vn}, BICs are states whose energy or
frequency lies within the continuous spectrum of radiation of an open quantum system
while remaining spatially localized. This localization can occur due to symmetry
constraints or through destructive interference processes \cite{Rev,Enginering}. By slightly varying
the system parameters that facilitate this decoupling, it is possible to weakly couple the
state to the continuum, leading to the formation of what is known as a quasi-BIC
\cite{Enginering,azzam2021photonic,Kang2023_NatRev_Phys}. The formation and manipulation of these states have become vital areas
of research, particularly in photonics and electronics, due to their applications in the
design of devices such as sensors, lasers, and filters \cite{Rev,10.3788/PI.2024.R01}.

Previous works devoted to study the relation between PC and BICs explore the role of 
flatbands \cite{PRL.132.173802_superBIC,PhysRevLett.114.245503}, topology \cite{PRL.113.257401_2014_topology},
subsymmetry \cite{ACS_pho_subsymm}, anisotropy \cite{ScientificRep2023_anisotropy},
Moire and twisted structures \cite{Qin2024_NatComm_moire,Qin2023_Light_twisted}, just to
name a few examples. Also, excellent reviews on the topic exists such as 
\cite{2023_ultrafastscience_rev,Koshelev2023_BIC_PV_rev,Kang2023_NatRev_Phys,2024_PhotIns_rev}. 

In this article, we explore the interaction between a one-dimensional periodic photonic
crystal and the physical properties associated with the presence of BICs and quasi-BICs
arising from the unit cell of the system.
We demonstrate how quasi-BICs generate extended bands in reciprocal space and how
these bands exhibit topological transitions. According to the bulk-boundary
correspondence, these transitions give rise to protected edge states that retain the
characteristic features of the original BICs. We conduct a careful analysis of the
symmetry properties underlying this nontrivial behavior through the lens of group
theory. Additionally, we reveal a regime characterized by flat bands within this
system, and we also present a symmetry analysis pertinent to this case.

The paper is organized as follows. In Section \ref{S1_model}, we present the
analyzed PC models along with the symmetry considerations that will be used in the
subsequent sections. Section \ref{S2_topology} focuses on the analysis of the
topological properties, beginning with a band perspective, and includes a
characterization of the Zak phase as well as the flat band regime. To validate the
nontrivial effects observed, we explore the consequences of the bulk-boundary
correspondence in Section \ref{S3_model}. We conclude the main part of the paper
with a discussion in Section \ref{S4_close}. Additional information, including the
standard computation of the Zak phase and a description of the initial bound states
and quasi-bound states used in this work, can be found in the Appendix and the
Supplementary Material (SM), %[*],
respectively.

\begin{figure*}[ht!]
\includegraphics[width=1.7\columnwidth,clip]{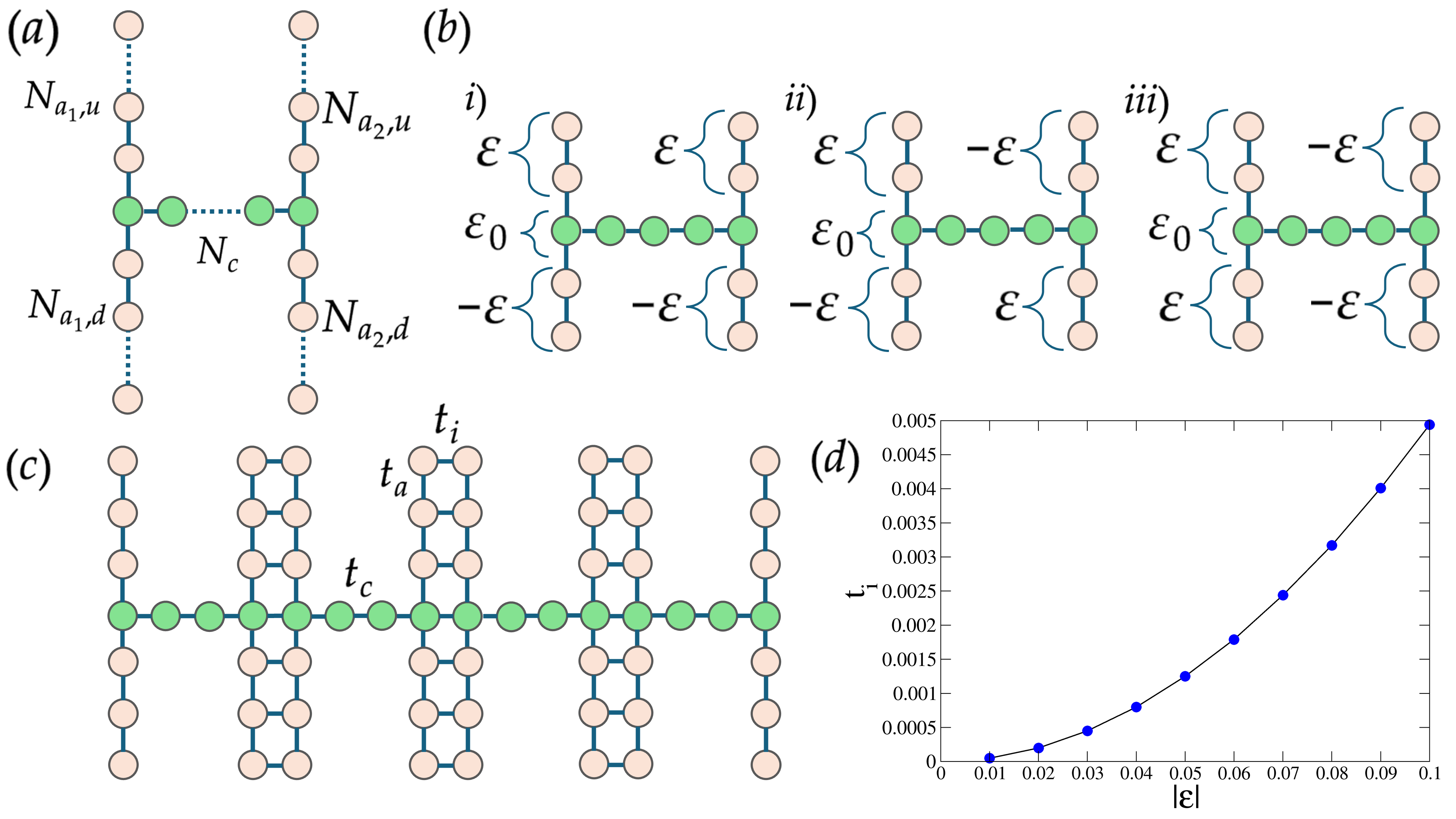}
\caption{(a) Unit cell configuration for the formation of a
generic one-dimensional crystal with ladder shape.   
(b) Configurations of the onsite energies corresponding to 
each of the symmetry groups presented in this work. 
i) $P2'm'm$, ii) $P2m'm'$ and iii) $P2'mm'$.  
(c) Sample of the periodic crystal showing the ladder
structures and the graphical definition of the hopping amplitudes.
(d) Graphical representation of the dependence of the critical value
of the intercell hopping ($t_i$) that produces the topological
band inversion versus the fixed value of the onsite detuning
magnitude $\vert \epsilon \vert$ for which this inversion happens.}
\label{fig1}
\end{figure*}

\section{Model: Structure, parameters and symmetries}\label{S1_model}

The structure of the crystal is the first piece of information needed to define
the physical model. The configuration in real space is described by a unit cell,
represented in FIG. \ref{fig1}.a. Generally, this unit cell is composed of a central
finite chain containing $N_c$ sites, with four arms attached: the upper left arm, which
has $N_{a_1,u}$ sites; the lower left arm, which has $N_{a_1,d}$ sites; the upper right
arm, which has $N_{a_2,u}$ sites; and the lower right arm, which has $N_{a_2,d}$ sites.

We consider only first nearest-neighbor interactions. However, several couplings can be
defined, which can, in principle, be varied independently in the model. We start by
defining the hopping interactions. First, we look at the couplings between the sites
within the central chain, denoted as $t_c$. Next, we consider the couplings between the
sites within each arm. Although these couplings can be defined separately for each arm,
we will use a single parameter to represent all hoppings, denoted as $t_a$. This
simplification allows us to observe the topological effects even in this basic model.
Lastly, we need to include a hopping parameter that connects the unit cells,
referred to as inter-cell hopping, which will be denoted as $t_i$.

To explore various types of symmetries in our crystal, we define onsite energies
separately. We assign an onsite energy for the central chain, denoted as
$\varepsilon_0$, and a specific onsite energy for each arm, denoted as
$\varepsilon_{a_{i,j}}$. Here, $i \in \{1, 2\}$ represents the left ($i=1$) and right
($i=2$) arms, while $j \in \{u, d\}$ indicates the upper ($j=u$) and lower ($j=d$)
arms. This setup completes the definition of the general features of the unit cell.

As this work focuses on studying the topological properties of a periodic structure, we
construct a one-dimensional crystal based on this unit cell by defining a
one-dimensional lattice vector that extends horizontally. A portion of the resulting
crystal structure, along with the definition of the hopping terms, is illustrated in
FIG \ref{fig1}.c, which also includes a graphical representation of the hopping terms.

To investigate the relationship between crystalline and global symmetries, we will
examine specific connections among the onsite energies. First, we will establish a
reference zero energy, which we define as the onsite energy of the sites in the chain,
denoted by $\varepsilon_0 = 0$. With this reference point, we can express the other
onsite energies as being \textit{detuned} from this baseline energy.

In the following, we will examine three configurations. To simplify our analysis, we
will maintain an equal number of sites for all arms, represented by the generic index
$N_a$. To classify the real-space periodic configurations, we will utilize group
theory, specifically Frieze groups \cite{Frieze_gr1}. These are discrete groups that
describe periodic one-dimensional structures where all sites reside in the same
plane, which we will consider to be the $x-y$ plane.

Without considering the onsite energy configuration and following the notation used
in standard crystallographic tables \cite{cryst_table}, the real-space symmetry of our
structure is characterized by the Frieze group $P2mm$. This group consists of a mirror
reflection with respect to the $x$ axis, denoted as $M_x$; a mirror reflection with
respect to the $y$ axis, $M_y$; and spatial inversion relative to the center of the
unit cell, $I$. In this context, inversion and a twofold rotation about an axis
perpendicular to the plane of the structure are equivalent, which accounts for the
use of the numeral 2 in the group notation. It is important to note that the first
$m$ in $P2mm $ corresponds to the $ M_x $ operation, while the last $m$ corresponds
to the $M_y$ operation. In the following, our primary focus is to explore how the
incorporation of different onsite symmetries can lead to nontrivial topological
behavior.

To achieve this, we will define the onsite energies at different arms in a manner that
creates an effective two-color symmetry within the system. This can be accomplished by
assigning negative and positive energies of equal magnitude. The three configurations
examined in this study, based on these constraints, are presented in
FIG. \ref{fig1}. b.

We observe that the inclusion of onsite energy introduces an additional degree of
freedom that must be incorporated into the complete symmetry description of the
crystal. Since there are only two possible values for the onsite energies of the
arms, we can introduce a symmetry operation, $C$, which transforms a site with energy
$\varepsilon$ into a site with energy $-\varepsilon$. This is the emergent two-color
symmetry that we combine with the spatial group. This operation alone does not
constitute a symmetry for the configurations of interest. Additionally, not all
transformations within the original $P2mm$  group will serve as symmetries of the
system. However, by combining the global symmetry with the spatial operations
$X$ from the $P2mm$ group, the structure will become invariant under combined
symmetries represented as $CX$.

This led us to introduce extensions of the parent Frieze group that incorporate
operations with the global symmetry $C$. This is similar to the situation in magnetic
space groups, where time-reversal symmetry is broken, resulting in a two-color
symmetry that links two different spin states. The description of these magnetic
structures is achieved through so-called two-color groups \cite{Bradley_2010}.

We follow the same procedure and extend the parent group $P2mm$ by defining the
combined two-color and spatial symmetries: $2' = CI$, $M'_x = CM_x $, and
$M'_y = CM_y$. This allows us to classify the configurations shown in FIG.\ref{fig1}.b
using the \textit{two-color Frieze groups}: $P2'm'm$ for case i), $P2m'm'$ for case
ii), and $P2'mm'$ for case iii). This notation clearly indicates which symmetries are
combined and which are purely spatial. A straightforward conclusion we can draw from
this is that the systems of interest in this work exhibit partial $C$ symmetry
breaking. In the following sections, we will explore the consequences of this broken
symmetry concerning the band topology of the periodic structures.

\section{Topological bands induced from bound states}\label{S2_topology}

To discuss the topological effects, we will begin by analyzing the states that arise from
the finite unit cell. In this case, the Hamiltonian can be represented as a finite matrix,
which can be diagonalized directly in real space to obtain the spectrum. Previous studies
have shown that this finite geometry supports BICs when connected in a two-terminal transport configuration \cite{sofia}. The
signatures of these bound states can be identified by computing the local density of
states (LDOS) within the energy range of interest.

Additionally, the location of the bound states in energy is influenced by the parity of
the number of sites in both the chain and the arms. We can encounter four cases, which
can be compactly denoted as ($N_a$ = Even/Odd, $N_c$ = Even/Odd). By combining this
parity property with the types of symmetry groups, we arrive at several distinct cases.
A summary of the types of BICs and quasi-BICs for each combination is presented in the
Supplementary Material (SM).

Although we have outlined all these cases, the topological features we aim to report
do not depend significantly on the parity of the number of sites. Therefore, we will focus on one specific configuration, namely ($N_a$ = odd, $N_c$ = even). This choice is made because this configuration exhibits two bands near zero energy. If necessary, we will briefly comment on any pertinent differences with the other cases.

\begin{figure}[ht]
\includegraphics[width=\columnwidth,clip]{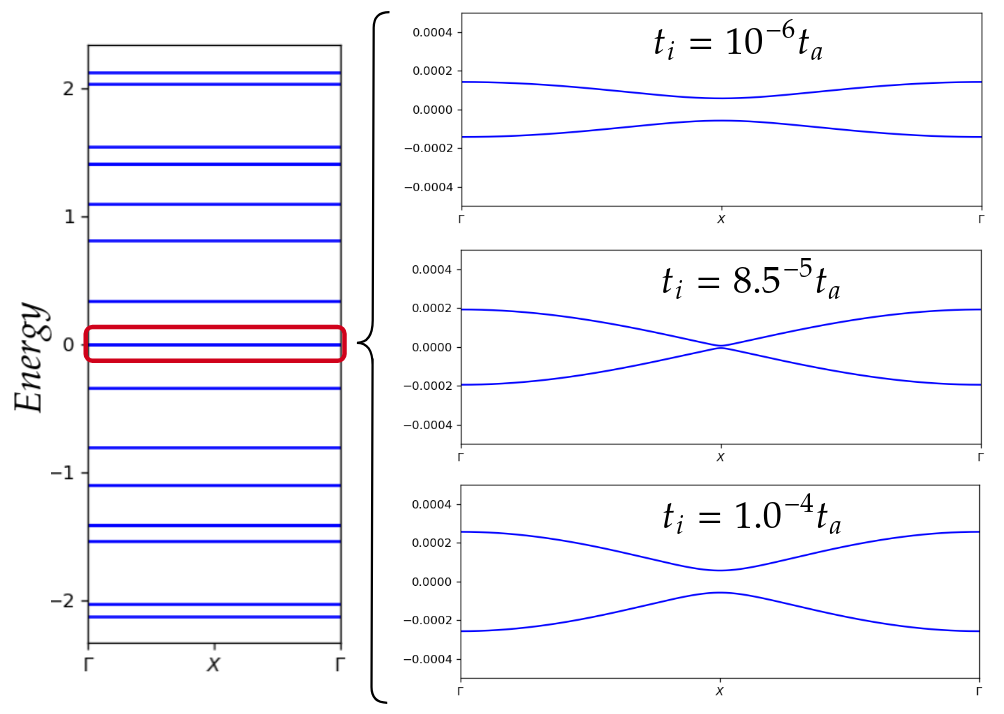}
\caption{Left: Band structure for a representative case of the 
$P2'm'm$ group, using as parameters $t_a=t_c=1$, $\vert\epsilon|=0.01$.
Right: Zoom in the central pair of bands induced from bound states, showing
the evolution with respect to $t_i$ to spot the band inversion associated to the
topological transition.}
\label{fig2}
\end{figure}

\subsection{Band inversion mechanism}
Once the periodic structure is established, we can obtain bands induced by the quasi-BICs
of the finite unit cell. The parameters that will be relevant for the subsequent
discussion are the magnitude of the detuning energy, denoted as $|\varepsilon|$, and the
amplitude of intercell hopping, referred to as $ t_i $. For our analysis, we will set the
other hopping amplitudes to one, specifically $ t_a = t_c = 1$. Consequently, all
quantities will be expressed in units of $t_a$ and $t_c$. This choice will be
maintained throughout our analysis, along with the condition $ \varepsilon_0 = 0 $.

To begin studying the bands, we numerically solve a Schrodinger equation with a
suitable Hamiltonian based on the previously described unit cell, valid for the
PCs under the paraxial approximation \cite{quantum-optical}. In first place, we
explore the effect of intercell hopping under the condition of a small detuning,
$|\varepsilon| = 0.01$.
Initially, we will set a very low value for $t_i$ and observe that the bands
exhibit an atomic-like behavior, closely resembling the finite BICs of the unit
cell spectrum. This initial case is illustrated in FIG. \ref{fig2} for a specific
example of $(N_a, N_c) = (3, 4)$.

The advantage of selecting an (odd, even) configuration is that a pair of bands
derived from the quasi-BICs appears near zero energy, allowing us to investigate the
band topology of this set as a representative subset of the other bands induced by
quasi-BICs. The evolution of this band pair concerning $ t_i $ is shown in the left
panel of FIG \ref{fig2}.b, where a clear band inversion is identified. As is well
known, this indicates a topological transition \cite{RevModPhys.88.021004}. 
Interestingly, these bands are very close in energy, which accounts for the
narrow range in which the band inversion occurs.

As $ t_i$ increases beyond the value that produces the band inversion, the energy gap
between the bands increases monotonically. This is a desirable feature, as it allows
for the control of the gap size as a function of intercell hopping. This monotonic
dependence continues until the inverted bands intersect with adjacent bands above
and below, leading to new band inversions, which can alter the topological character
of the bands. The properties described above are observed in all other parity
configurations for $N_a$ and $N_c$, with the only difference being the specific
energy locations of the quasi-BIC-induced bands.

In the second stage of our analysis, we examine the influence of other
parameters in the model concerning the band inversion effect. Our numerical
calculations reveal that the hopping amplitude $t_c$ does not significantly
affect the control of band inversion. However, we found that the hopping
amplitude $t_a$ can indeed induce band inversion, but this is contingent upon
the magnitude of the onsite detuning $|\varepsilon|$. Specifically, the value
of $t_a$ required to achieve band inversion increases as $|\varepsilon|$ becomes
larger. This behavior is analogous to the relationship between $|\varepsilon|$
and $ t_i$. Consequently, we will concentrate on the pair ($|\varepsilon|$,$t_i$)
for now on and will postpone the exploration of the interplay between all three
parameters for future research. 

In summary, the most significant effect of ($|\varepsilon|$) is its ability to
directly control the size of the gap between the bands. 
As a result, there exists a competition between $|\varepsilon|$ and
$t_i$ (when $t_a$ is held constant), meaning that the value of $t_i$ necessary
for inducing inversion increases with higher $|\varepsilon|$. We provide a
graphical representation of this functional dependence for fixed $t_a$ and
$t_c$ in FIG. \ref{fig1}.d.

\begin{figure*}[ht]
\includegraphics[width=2.0\columnwidth,clip]{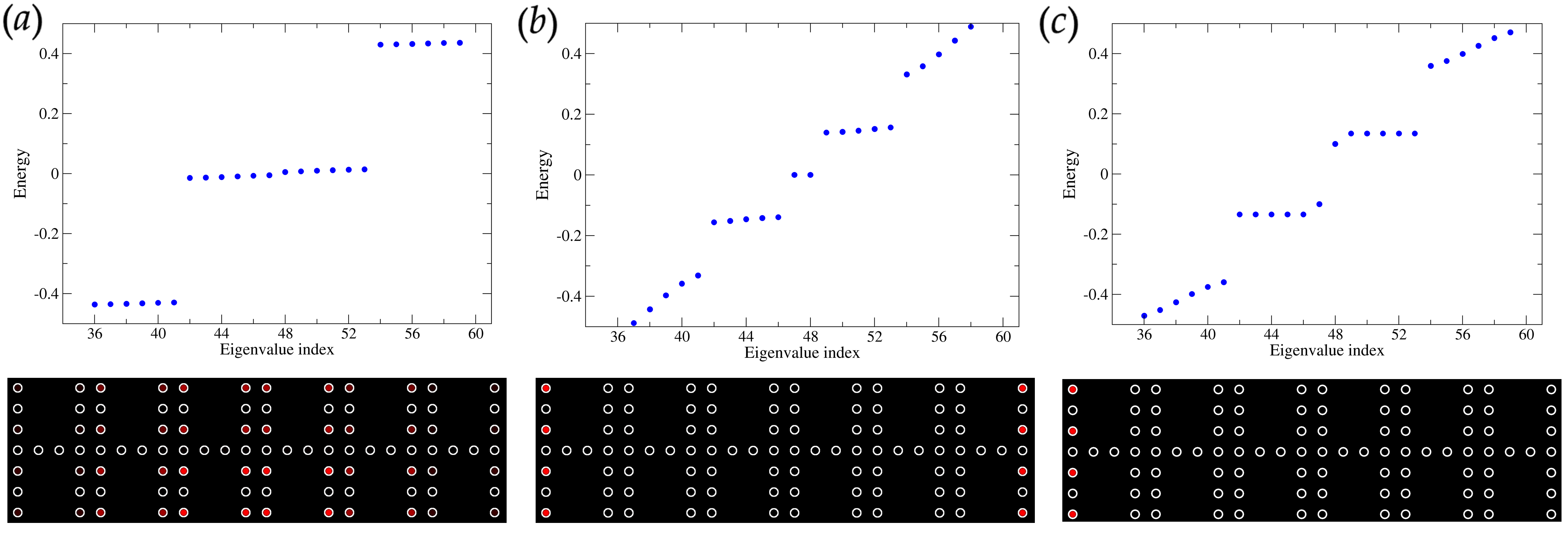}
\caption{(a) Top panel: Eigenvalue spectrum for a finite system with six unit cells
in the trivial phase of $P2'm'm$ group. Bottom panel: LDOS for the finite system
in the trivial phase of $P2'm'm$ group. 
The value of intercell parameter is $t_i=0.0015$.  
(b) Top panel: Eigenvalue spectrum for a finite system with six unit cells
in the topological phase of $P2'm'm$ group. Bottom panel: LDOS for the finite system
in the trivial phase of $P2'm'm$ group.
The value of intercell parameter is $t_i=0.15$.
(c) Top panel: Eigenvalue spectrum for a finite system with six unit cells
with flat bands for the $P2'mm'$ group. Bottom panel: LDOS for the finite system
with flat bands for $P2'mm'$ group.
The value of intercell parameter is $t_i=0.09$.
For all the above calculations, the other parameters where set to: $t_a=t_c=1$ and 
$\vert \epsilon \vert = 0.1$.
For all top panels: red circle indicates the states for which the LDOS is computed.
For all bottom panels: The intensity of the sites indicates 
the magnitude of the modulus of the wavefunction.  
}
\label{fig3}
\end{figure*}

\subsection{Zak phase characterization}

To support the band inversion mechanism discussed earlier, we compute a topological
bulk invariant. For a periodic system, the Zak phase \(\phi\), which corresponds
to the Berry phase defined in reciprocal space \cite{zak_PRL1989}, is a natural
choice. The procedure for numerically computing this invariant for one-dimensional
structures is detailed in the Appendix. Because there are no degeneracies between
the bands, we can compute the Berry phase for each band separately. This phase is
well-defined up to a multiple of \(2\pi\) and encodes information about the
evolution of the wavefunction in reciprocal space.

In general, the Zak phase is not quantized and can take any value within the
interval \([- \pi, \pi]\). However, when spatial symmetries such as spatial
inversion or mirror reflections are present, the Zak phase can take quantized
values that remain robust under symmetry-preserving perturbations
\cite{vanderbilt_book}.

In a one-dimensional periodic system, the quantized Zak phase can take on two
distinct values: $0$ and $\pi$ \cite{Inv_TI_PRB2011}. These values
conventionally represent the trivial and nontrivial phases, respectively. 
Depending on how the unit cell is defined, the calculated values of $\phi$
may differ from these quantized values. However, they can always be transformed
to match the quantized values by appropriately selecting the origin in real
space.

In the following, we will examine the topology of the bands in general terms
to analyze how the symmetry group of the structure influences this topology.
We begin with the case of $P2'm'm$. Utilizing the numerical procedure
(see Appendix), we compute the values of \( \phi \) for various configurations
of $N_a$ and $N_c$, using the unit cell origin choice depicted in Figure
\ref{fig1}.a. Our findings indicate that $\phi$ takes on quantized values that
are independent of the parity of $N_a$ band $N_c$.

\begin{align}
     \phi_I &= \pi-\frac{\pi}{N_c}, \label{zak1}\\
     \phi_{II} &= 2\pi-\frac{\pi}{N_c}. \label{zak2}
\end{align}
o dependence on $N_a$ was found. As noted, $\phi_I$ and $\phi_{II}$ differ by
$\pi$, allowing them to distinguish between the trivial and nontrivial phases.
When we repeat the calculation for the $P2m'm'$ group, we find the same
result: $\phi$ also takes quantized values, represented by equations
(\ref{zak1}) and (\ref{zak2}). In contrast, for the $P2'mm'$group, all
configurations exhibit a non-quantized $\phi$.

The behavior of these three cases can be understood in a unified manner by
considering the conditions required to protect a quantized Zak phase.
Generally, a Zak phase needs spatial inversion and/or mirror reflection along
the periodic direction to maintain its quantized values. By examining the
groups involved in this study, we can identify this rule. Specifically,
$P2'm'm$ has $M_y$ as the protecting symmetry, while $P2m'm' $ relies on
inversion as the protecting transformation. In contrast, $P2'mm' $ possesses
$M_x $ as a purely spatial operation, which is not aligned with the periodic
direction. Since the other two symmetries are not purely spatial, they
cannot safeguard a quantized value for $\phi$.

Once we have characterized the topological phases, we can examine their
robustness by varying the model's parameters, ensuring that we do not break
any of the protecting symmetries. We have conducted numerical simulations that
confirm the value of the Zak phase remains robust against perturbations, as
long as gap closings do not occur during the process. This provides further
evidence that the topological transitions and the associated topological
phases are accurately represented by the value of the Zak phase.

\subsection{Flat bands induced from mirror anti-symmetric bound
states}\label{flatband}

Despite having a non-quantized Zak phase, the $P2'mm$ configuration has a
band structure characterized with flat bands. Representative band structures
illustrating this flat band behavior are presented in the supplementary
material (SM).
The presence of these flat bands arises from the formation of two types of
states, which is a direct consequence of the mirror symmetry $M_x$ in this
group. Since $M_x$ commutes with the Hamiltonian $H$, the eigenstates of
$H$ can be classified according to the eigenvalues of $M_x$. This results
in the existence of mirror symmetric and mirror anti-symmetric states.
As observed in the spatial structure, the mirror symmetry has a specific
effect on the arm states, as they get interchanged by the operation.
Therefore, we can further classify the states into arm-localized and
chain-localized states. The significance of this classification lies in
the fact that the chain-localized states are always symmetric with
respect to $M_x$. In this arrangement, the symmetric arm states tend
to hybridize with the symmetric states of the chain, leading to
typically dispersive bands. In contrast, the anti-symmetric states of
the arms do not couple to the states in the central chain, resulting
in non-dispersive bands across the momentum space. This explanation
clarifies the origin of the flat bands, which are induced by mirror
anti-symmetric bound states.

Furthermore, we can understand the absence of flat bands in the other
two symmetry groups: since there is no pure spatial symmetry in the
transverse direction (perpendicular to the periodic direction) in
these structures, there is no possibility for the formation of
anti-symmetric states that can be decoupled. Consequently, all states
in those systems become dispersive.

\section{Bulk-boundary correspondence in finite systems}\label{S3_model}

To fully characterize the nontrivial nature of topological bands, the
next step involves examining the realization of bulk-boundary correspondence
within a finite geometry. This is where the choice of the unit cell proves
to be advantageous, as we can construct a suitable finite system by
repeating the unit cell a whole number of times.

In this scenario, the emergence of a correspondence is indicated by the
presence of protected edge states within the nontrivial gap. To
investigate this, we compute the spectrum for a finite sample. For this
analysis, we use a larger detuning magnitude, $|\varepsilon| = 0.1$, to
more clearly resolve the in-gap states. The critical value of $ t_i$ for
the inversion to occur in this regime is approximately $0.049$.

We present the eigenvalue spectra for the $P2'm'm$ symmetry in both the
trivial and nontrivial phases for a finite system, as shown in the top
panels of FIG. \ref{fig3}.a and \ref{fig3}.b, respectively. In the
nontrivial case (FIG. \ref{fig3}.b), we can directly observe the
emergence of two degenerate in-gap states. These states are edge
states, protected by the nontrivial nature of the bulk bands, and
serve as a signature of the bulk-boundary correspondence.

To analyze the form and spatial localization of the states near zero
energy, we plot the LDOS for one of the in-gap states in the bottom panel
of FIG. \ref{fig3}.b. The edge localization is clearly demonstrated,
resembling a typical nontrivial edge state found in a SSH-like system
\cite{short_course_top}. For comparison, we also illustrate the LDOS for a
state with the same index in the finite system during the trivial phase,
shown in the bottom panel of FIG. \ref{fig3}.a. In this case, there are
no in-gap states, and the LDOS shows no signs of edge localization.

Although this feature is not protected, it is important to note that for the
$P2'mm'$ group, a finite geometry will reflect the bound states and the
associated flat band characteristics discussed in Section \ref{flatband}. In
FIG. \ref{fig3}.c, we present the eigenvalue spectrum and the local density
of states (LDOS) for a representative set of parameters. In this case, since
there is no quantized Zak phase, we observe that there are no protected
in-gap states. This is because the states that do appear in the gap are not
required to remain degenerate and can potentially hybridize with the states
from the flat bands. Interestingly, when these states are present in the
gap, their localization resembles that of a protected edge state, but only
show a one-sided localization in real space (see the bottom panel in
FIG. \ref{fig3}.c). This behavior differs from that of topological edge
states, which exhibit symmetric localization.

\section{Discussion}\label{S4_close}

In summary, we have designed a one-dimensional crystal that takes advantage
of BICs and quasi-BICs produced by the unit cell design, allowing us to
realize a unique type of band structure induced by these states. By
employing a combination of onsite and spatial symmetries, we establish
several extended symmetry groups that can describe various physical
scenarios for the band structure of the system.

By analyzing the parameter space, we achieved a topological transition
controlled by the intercell hopping parameter $t_i$, similar to the
well-known SSH chain. The topological character of the bands is characterized
by the Zak phase, and we identified two structures with nontrivial bands,
corresponding to the $P2'm'm$ and  $P2m'm'$ symmetry groups. This finding
is further supported by finite system calculations that reveal the emergence
of protected edge states, which indicate a bulk-boundary correspondence.

Additionally, for the structure with no quantized Zak phase ($P2'mm'$), we
observed flat bands in the spectrum. We traced the presence of these flat
bands to a symmetry-protected decoupling of the inducing states, which is
a result of the $M_x$ mirror operation.

%% agregar mención a comprobación que son BICs
Regarding the BIC nature of the topological edge states, the standard method
to investigate the system involves connecting it to a continuum of states
in a transport configuration. In our case, we employ two simple chain leads
with zero onsite energy and a single hopping amplitude, denoted as $t_a$, as
defined earlier. These leads are connected to the external sites of the chain.

In this configuration, we calculate the transmission function and the local
density of states (LDOS). Results presented in the SM indicate that the
topological edge states do not couple to the continuum since there is no
signal in the energy range where the nontrivial states are located. Additionally,
the LDOS with the connected leads clearly shows the presence of edge states at
the corresponding energy levels. These observations serve as clear evidence
that the topological states are indeed classified as topological BICs.

The phenomena described above make this family of one-dimensional crystals
a simple yet versatile platform for exploring the interaction between
topological phases and bound states in the continuum. This is especially
relevant because the transport properties of nontrivial states in finite
systems can be investigated experimentally, allowing for the evaluation of
their robustness and potential applications. Furthermore, the presence of
flat bands is a promising feature that can be used in finite configurations,
for example, to manipulate and adjust the localization of boundary states,
thanks to the inherent characteristics of these systems.

It is worth to note that there are several ways to extend the results of
this work. Firstly, the role of symmetry breaking can be further explored.
This can be combined with structural modifications, such as investigating
different types of onsite and hopping configurations. Additionally, the
characterization of the physical properties of finite systems can be
broadened by examining transport configurations, where the interactions
between topological, trivial, and flat band-induced states can be studied.
\\

\section*{Acknowledgments}
S.P. acknowledges financial support from ANID-Subdirección de Capital Humano/Doctorado Nacional/2025-21251220, DP USM PIIC N° 017/2025. S.B. acknowledges financial support from postdoctoral position by DGIIE USM. P.A.O. acknowledges financial support from DGIIE USM PI-LIR-24-10 and FONDECYT Grant No. 122070. S.P. and P.A.O acknowledge financial support from FONDECYT Grant No. 1230933.

\section*{Appendix: Zak phase calculation}\label{Append}
\renewcommand{\theequation}{A.\arabic{equation}} % Redefine equation numbering for Appendix A
\setcounter{equation}{0} % Reset equation counter for Appendix A

The numerical procedure that we used to compute the Zak phase (Berry phase
in reciprocal space) for the one-dimensional systems in this work are based
on a discretization of the expression for the Zak phase for a continuous
momentum parameter $k$. This continuous formulation is given for a
nondegenerate band with index $n$ by \cite{vanderbilt_book}
\begin{equation}
    \phi = \oint_{BZ}A_n(k)dk=\oint_{BZ}\langle u_nk|i\partial_k u_nk \rangle dk,
\end{equation}
where $|u_nk \rangle$ is the periodic part of the Bloch eigenstates of the
tight-binding Hamiltonian, where the gauge has been fixed such that, for 
the Brillouin zone ends we have that $|u_{nk=2\pi/a} \rangle=e^{-2i\pi x/a}
|u_{nk=0} \rangle$, with $a$ and $x$ being the lattice parameter and the
horizontal position of the unit cell sites. 

Discretization of this expression implies the construction of the set of numbers
\begin{equation}
    M^{k_i,k_{i+1}}_n = \langle u_{nk_i}|i\partial_k u_{nk_{i+1}} \rangle
\end{equation}
for each band separately. Here $k_i$ represents a set of sampling points along
the whole brilluoin zone which carry the discretization. 
Finally, the Zak phase is computed using \cite{vanderbilt_book}
\begin{equation}
    \phi = -\text{Im} \hspace{0.1cm } \ln \prod det(M^{k_i,k_{i+1}}_n),
\end{equation}
where a branch of the logarithm function must be selected to have a 
well-defined phase.

\bibliography{refs}% Produces the bibliography via BibTeX.

\end{document}